\documentstyle[preprint,aps,psfig]{revtex}
\begin {document}
\title{Alignment in hadronic interactions}

\author{Tadeusz Wibig}

\address{Experimental Physics Dept., University of \L \'{o}d\'{z}, \\
ul. Pomorska 149/153, PL-90-236 \L \'{o}d\'{z}, Poland}

\date{\today}

\draft

\maketitle

\begin{abstract}
The alignment of the products of very high energy interactions seen in mountain
altitude experiments is one of the most puzzling phenomena
in cosmic ray physics for quite a long time. The observations
of the Pamir and Chacaltaya emulsion chamber groups and by
the Tien-Shan extensive air shower experiment, together
with a very clear event seen in the Concorde French--Japanese
experiment in the stratosphere, makes the experimental basis very substantial.
In the present paper a novel possible explanation is put forward.
\end{abstract}


\section{Introduction}
The Pamir experiment comprises an X-ray film chamber which
contains events with  genuinely correlated
dark spots produced by very energetic cosmic ray shower particles
distributed implausibly asymmetrically\cite{pamir}. This effect,
called ``alignment'' occurs, according to Ref.~\cite{thre},
at a primary particle energy of $8 \times 10^{15}$eV above which the rate
increases rapidly with interaction energy.
For the energies of interest the cosmic ray
flux is very low, so the statistics in the
Pamir experiment are very limited; specifically only 62
events of visible electromagnetic energy in the range from 700 to 2000 TeV.
Problems with the statistics and the quite complicated methodology of the
large area X-ray film calorimeter make the measurements very difficult.
Thus, independent confirmations by the Chacaltaya group\cite{chacaltaya},
the Tien-Shan extensive air shower experiment\cite{tien}, and
the one event of high quality recorded in the Concorde French-Japanese
experiment\cite{concord}, are indispensable.
All this implies rather strongly that there is the
azimuthal asymmetry of particle production at energies above about $10^{16}$eV.

Some explanations exist in the literature already.
The most conservative one is the fluctuation explanation by
J.N. Capdevielle\cite{JNCinUtah}. Calculations show that with conventional
fluctuations of the elementary act the probability of producing the
Concorde event could exceeded the ``5$\sigma$'' level, the commonly accepted
limit for a ``new physics'' discovery. However, the increase
of the fraction of aligned events to (27$\pm$0.09)\% at the highest
energies seen, makes this explanation less probable \cite{thre}.
Even if it could be
responsible for the one stratospheric event, which is, in fact, slightly
different than the rest of ``X-ray aligned'' events, taken all together
a different solution is required.

The rotating nuclei fragmentation hypothesis of Erlykin and
Wolfendale\cite{AWWatGranSasso} is naturally connected with the
postulated increase of the fraction of heavy nuclei in the primary cosmic
ray flux around $10^{16}$eV (up to 50\%). However,
due to the existence of the few TeV threshold in the
X-ray film technique, the  experiments noticeably
favour primaries of higher energy per nucleon
(for the same energy per particle).
and the existence of heavy nuclei alone
is not enough to account for the 30\% alignment that is needed.
The problem was discussed in detail in Ref.~\cite{AWWatGranSasso}.
The mechanism of delayed
fragmentation of fast rotating nuclei need to be established
theoretically as well as the possibility of a reduction of its cross section
for interaction with air nuclei.

Additionally, the observation that aligned events are more abundant
in the vertical component, supports the concept that the origin of the
phenomenon is a deeply penetrating cosmic ray particle (most likely a
proton).

The original explanations by the Pamir group \cite{Mukhamedshin}
is supported by extensive calculations made by Mukhamedshin.
He shows that the Pamir data required a significant change of the
particle production act. Particle creation with a few GeV
transverse momentum in a one plane cascade-like process is needed but even
this is not enough to explain the data. Perhaps there is,
additionally, a heavy, long lived and relatively weak
interacting particle? Some theoretical ideas of the ``new
physics'' were discussed in, e.g., Ref.~\cite{royzen}.

\section{The proposed concept of alignment in the
hadronization mechanism}

The inelastic collision of two elementary particles
is conventionally treated as a
scattering of particle constituents, i.e. partons, associated with
the creation of excited, intermediate
objects (jets, strings, chains, or fireballs), followed by
their hadronization.
Such a picture is confirmed by Bose-Einstein correlation
studies, where the time and spatial extensions of the newly created
particle source is seen and measured.
Both the collision and excitation processes can not be fully and
quantitatively described by QCD.
Specifically the hadronization is, by definition, a low momentum transfer
phenomenon and at present it can be taken into account only using
models.
Most of the effects seen in cosmic ray interaction physics are
low $p_t$ effects. Alignment seems to be an exception, but it will
be shown later that it is probably not.

It is well known that
the scattering can be described in the impact parameter representation.
This point is used to maintain unitarity by the eikonal formalism,
both by the dual parton \cite{DPM} and by relativistic string
\cite{LUNDb} models. For our objective, however, the details are not crucial.
Differences between particular model realizations
such as, for example parton scattering cross sections (at given impact parameter $b$)
or the character of the intermediate object created, do not change
the essence of the concept of this paper, and they will not be
discussed here.
The implications of the impact parameter picture considered strictly
are valid for most of the models of multiparticle production.

The important characteristic of the
string (this name will be used hereafter to label the intermediate excited
objects in spite of their fireball, chain, or jet nature)
is its mass. It is certainly proportional to the interaction
energy [available in the center of mass system (c.m.s.) --
$\sqrt{s}$], but it could also
depend on other collision parameters such as the impact parameter $b$.

On the other hand, the string mass could be the random variable and it could
fluctuate from collision to collision according to the respective probability
distribution. The most familiar way of considering the chain mass behaviour
is to use the parton distribution functions $F(x,Q^2)$ which describe the
probability density of the colliding parton for a fraction $x$ of the
total hadron momentum ($Q^2$ defines the scale).
The particular shape of $F$ (with its $s$ dependence) could not be obtained
from QCD. However, recent progress in lepton scattering at HERA have expanded
our experimental knowledge of $F$ significantly \cite{hera}.

If we denote by $b$ the impact parameter of the inelastic collision
of two hadrons (protons) and by $M$ the masses of two strings created
(for simplicity we can take the masses to be equal, but this is
not necessary) then, due to the conservation laws,
the strings carry an angular momentum of
\begin {equation}
J_3 ~ \approx ~b c \: \sqrt{s}\: \left( {M \over \sqrt{s}} \right)^2~.
\label{dl}
\end {equation}
Here, $M$ is, to be precise, the energy of the (rotating) string
in its c.m.s.. It is of the
same order as the energy (mass) of the string in the co-rotating frame.

The angular momentum of the string is related to its (end) rotation velocity,
$\omega$. In the case of
particular model of the string (Nambu-Goto-Polyakov, NGP)
\cite{ngoto}, which will be used later as a numerical example,
it is given by
\begin{equation}
J_3~=~{a\: R   \over 2\: \omega} \left[\: { { {\rm arcsin} 
\left( {\omega \, R / c} \right) \over  {\omega \, R / c} }
~-~\sqrt{1~-~\left( {\omega \, R  \over c }\right)^2} } \:\right]
~\approx ~{M \: c^2 \over  2\: \omega }~.
\label{ngp}
\end{equation}

Comparing Eqs.(\ref{dl}) and (\ref{ngp}), the angular
velocity of the relativistic string is of the order of
\begin{equation}
{\omega}~\approx~{c \over b} ~{ \sqrt{s}\over M}~.
\label{om}
\end{equation}

The above relations certainly does not hold for central collisions,
and also in all other cases it should be treated rather approximately, just to
illustrate the importance of relativistic rotation of fragmenting strings.

\section{The angular momentum problem of string fragmentation.}

\begin{figure}
\begin{center}
\begin{minipage}{75mm}
\psfig{file=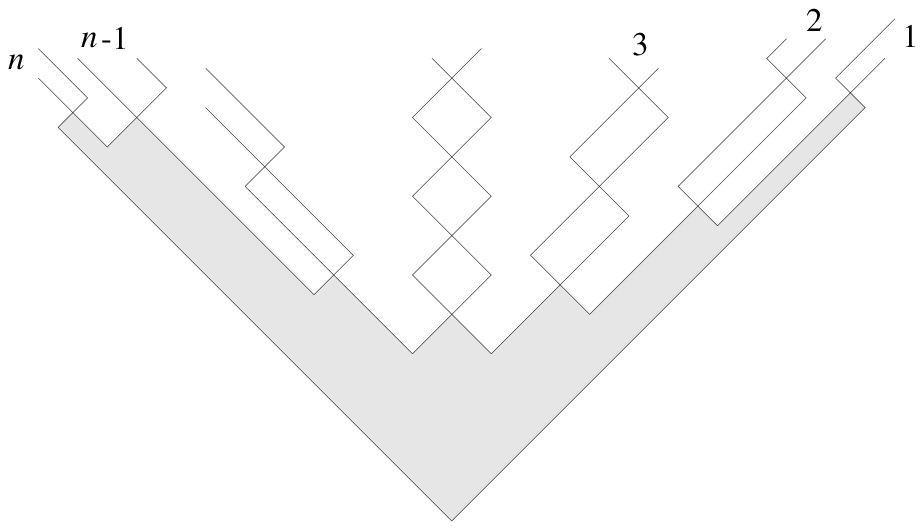,width=12cm}
\end{minipage}
\hspace{-11.25cm}
\begin{minipage}{97.5mm}
\vspace{.75cm}
\psfig{file=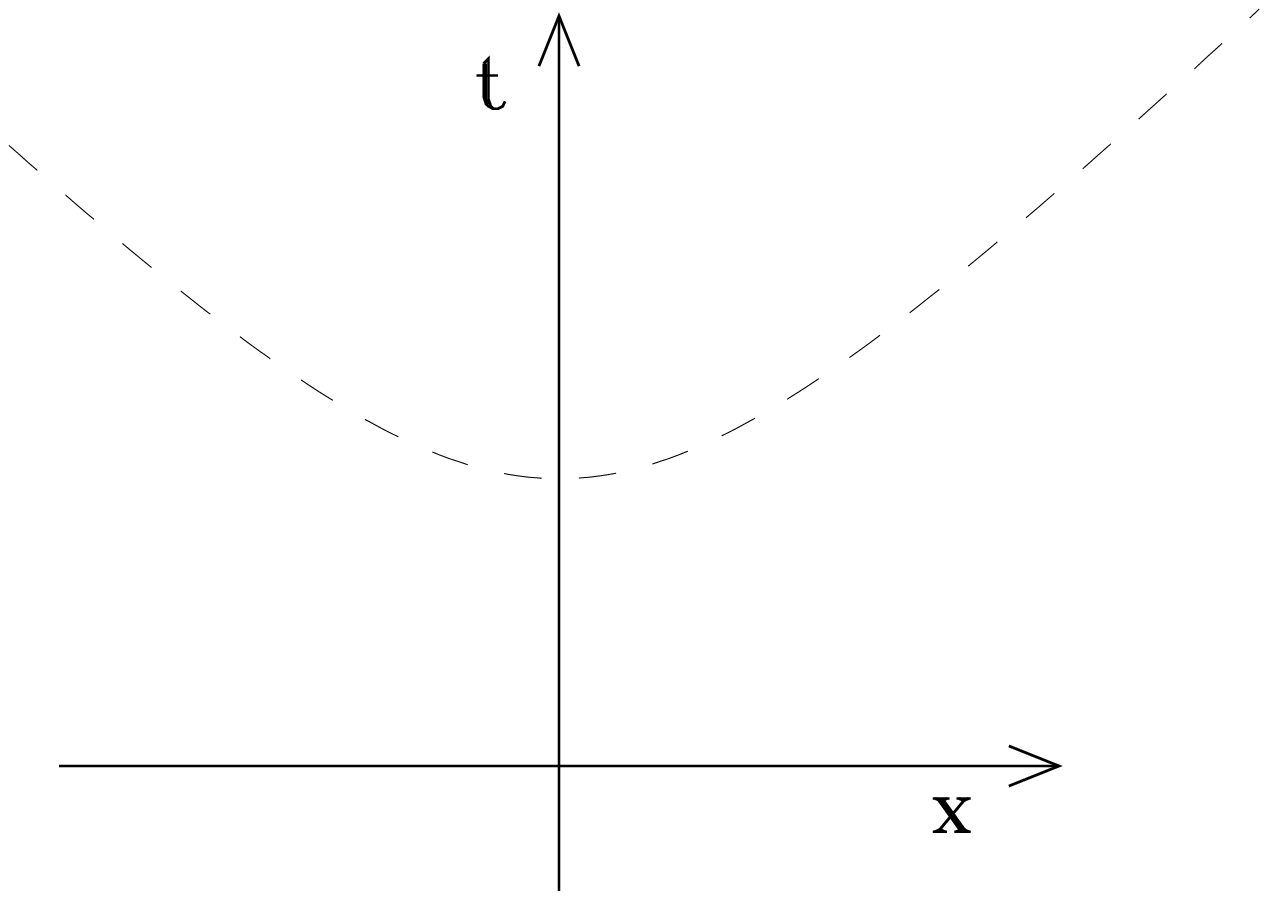,width=9.75cm}
\vspace{-.75cm}
\end{minipage}
\end{center}
\caption{
Space-time structure of the relativistic string fragmentation model.
\label{hadr}}
\end{figure}

If we have a (rotating) string of mass $M$, the next step to be considered
is its fragmentation.
The commonly used hadronization models can be cluster type-like
(HERWIG\cite{herwig}) or string fragmentation-like (LUND\cite{LUND}).
Nevertheless, both deal with a one-dimensional coloured field structure, and
this one-dimensionality is an important part of the models. The LUND
picture  possesses a well described
space--time particle production scheme. It is shown in Fig.~\ref{hadr}.
Details and particular model parameter values are adjusted to the measurements.
The recent and most accurate tests of hadronization models are made with
the precise data on the Z$^0$ energy from LEP\cite{lep}. 
The important point is
that just as in the case of $e^+$$e^-$ annihilation the linear structure of
the Z$^0$ decay created chain is (can be) very well justified
due to the vanishing of the string angular momentum. The fact that the same
procedures can be used for hadronization of fast rotating strings from
hadronic inelastic collisions is rather astonishing. However,
a closer look at particular Monte Carlo realizations exhibits many
modifications which make the previous amazement less surprising.

The problem with the angular momentum, or rather the lack of it,
seems to have a very long history.
The famous Fermi\cite{fermi} statistical model was published just
fifty years ago. It is interesting that the importance of the problem
was clear to Fermi. He described the exact way of avoid it by using the
impact parameter formalism. He even made some calculations, concluding that
{\em ``It was found in most cases that the results so obtained differ only
by small numerical factors from those obtained by neglecting the conservation
of angular momentum. This has been done as a rule in order to simplify
the mathematics.''} This is not very surprising taking into account the low
energies which Fermi had to deal with (in the main part of his paper);
these were of the order of hundreds of MeV to a GeV or so. However, an
interesting remark
can be found in the second last page where he discussed the {\em
``collisions of extremely high energy''}
(10$^{12 \div 13}$eV). He found that the conservation
of angular momentum reduces the produced pion multiplicity, and also
{\em ''...has the effect that the angular distributions of particles produced
is no longer isotropical...''}.

Ten years after Fermi's paper, when it become clear that the products
of high energy collisions are strongly collimated along the interaction
axis, Hagedorn published paper \cite{hage1} concerning the statistical
treatment of the {\em not-so-isotropic} angular (in c.m.s.) distribution of
collision products. In this paper interesting statements appeared: {\em ``This
whole question {\rm [the angular momentum conservation problem]},
though of practical importance, seems to be still not understood.
At least angular momentum conservation does not play an
important quantitative role. ... So at present it seems most reasonable
to disregard angular momentum at all,...''}. Although the model discussed
in the mentioned paper was assigned for central collisions, everything
was entirely correct. The problem of the most frequent, peripheral collisions,
however, remains.
Hagedorn himself, again ten years later, gave the solution in his
paper called ``The Thermodynamical Model''\cite{hage2}. The solution
was transient rather than fundamental. He introduced {\em the velocity
weight functions} which describes a part of strong interaction physics
(unknown from first principles) and {\em ``To this one should add
over-all conservation of angular momentum''}. And this temporary solution
survived thirty years! It can be found in more or less sophisticated
transcription in many contemporary string hadronization models.

Coming back to the LUND hadronization and its space--time structure
shown in Fig.~\ref{hadr}, an important remark has to be made concerning the
time sequence of the chain breakups. The dashed hyperbolic curve in the figure
represents the string point of the same proper time in their local co-moving
frames. This can be associated with the hadronization time of the string,
but, as is seen, in the string center of mass system hadrons occur
at different times. It is clear in Fig.~\ref{hadr} that the
slowest hadrons appear first,
while those with high velocities
(in the string c.m.s.), especially those containing initial string creating
partons, materialize last. This somehow puzzling
statement has, in fact, been known from the very beginning of
relativistic string theory (see e.g., Refs.~\cite{fff}). For the
relatively low energy collisions with only a few particles created it may
cause problems \cite{bo}.
To be precise, one has to note that, of course, in some cases,
 due to the random
nature of the process (which is slightly more complicated than that which
is shown
in the discussed simple graph), some exceptions can be expected.

The main point is that the central
part of the string fragments after some ``freeze-out'' time (about
1 fm/c) and the very end needs longer time (in the string c.m.s.) and we
can expect that the rotation speed could be large enough to bend the
particle production direction away from the interaction axis in the intervening
time. Of course
all particles in between will tend to lie on the one plane defined by the
impact parameter vector and in this way the particles created are apparently aligned in the laboratory frame
of reference. What should be noticed here is that the production of
particles with relatively high transverse momentum (with respect to the
interaction axis) is not due to any special high momentum transfer
process (new physics) but is simple a result of kinematic with the usual
non-perturbative hadronization.

\section{The curved string fragmentation.}
The consequences of the string rotation presented above
could, however, be quite wrong, because of the
well known fact of the non-existence of ``rigid'' rotation in highly
relativistic system.
The one dimensional fragmentation structure of the LUND model
should be extended making the problem definitely much more complex
when we deal with a fast rotating QCD string.

\begin{figure}
\centerline{\psfig{file=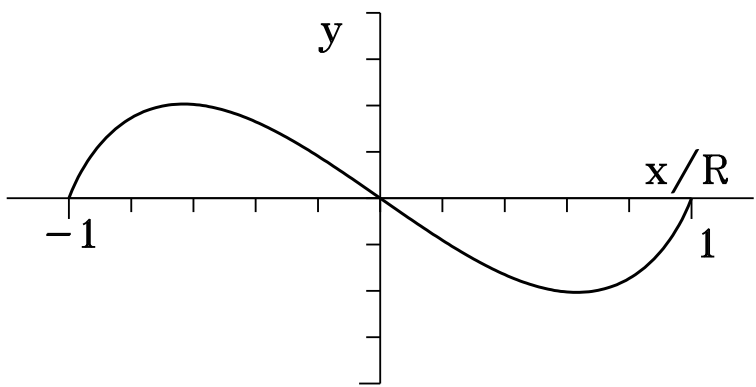,width=10cm}}
\vspace{.8cm}
\caption{ The shape of relativistic Nambu-Goto-Polyakov string.
The scale of distortion is arbitrary (see text).
\label{shape}}
\end{figure}

The main question here is the shape of the string. Detailed
calculations (see, e.g., Ref.\cite{ngoto}) assuming a particular model
of action for the string system show that there are some deviations
from the straight string shape. In Fig.~\ref{shape} the solution is
presented. The measure of the curvature of the string is in this case
proportional to the, e.g., rate of longitudinal expansion, so the vertical
scale in the figure is in this sense arbitrary. The solution
was obtained (in an analytical form) assuming small string
deflections by perturbation
of the straight string solution, thus, dropping some terms in the
general string evolution equation, which do not have to be necessarily
small in our case. The lack of an exact solution makes further examination
somehow uncertain, but we do expect that the general behaviour of
the real string shape is similar to the one shown in Fig.~\ref{shape}.

The main difference between the straight rotating string and the ``real''
(Fig.~\ref{shape}) string is that the end quark of the string (leading one)
still moves almost along the interaction axis direction while the inner part
of the string bends.

The problem of hadronization and specially its space-time structure
(similar to the one presented in the Fig.~\ref{hadr}), of the curved string
is, in general, unsolved.
Inertial forces acting along the string could have an effect on the
string area law
thereby changing the string ``decay'' constant.
Additionally, the problem of clock synchronization
becomes non-trivial for rotating frames.
Thus, if we can expected that the rotating string
should decay {\em later} the meaning of the word {\em later}, is not
entirely and indisputably defined.
This makes our further analysis
more uncertain, but nevertheless we will try to obtain some qualitative
results.

The time evolution of the curved NGP string (such as that shown in
Fig.~\ref{shape}) can be simply described by the symmetrical
expansion in both $x$ and $y$ directions as the
string length grows. If, at some instant, the string begins to break
(starting from its central part) the created particles will conserve
the expansion speed of the particular piece of the string. From
the figure it is straightforward that some of the particles created
at the beginning will follow the same direction given by the string
deviation. The momentum transverse to the interaction axis of
subsequent particles will be getting larger (it is proportional to their
longitudinal momentum) up to the moment when the
fragmentation of the curved part of the string begin. Further emission
angles will be smaller and smaller and in the end the rest of the
created particles will follow the leading quark direction.

The relatively slow growth of multiplicity (a power-law in $s$ with the
power index of 0.3 to 0.1 or quadratic in logarithm of $s$]
leads to the specific
scaling of the arrangement of particle creation points on the curved
string with the interaction energy.

As has been mentioned, the central part of the string gives, in the laboratory
system, one collimated jet. It, together with the very forward produced
particles forming another jet, leads to the clear binocular event
reported in Ref.~\cite{binoc} by the
Chacaltaya experiment.
The central and the forward jets are formed by many constituting hadrons,
thus they carry enough energy (and particles) to be visible
in the X-ray chamber as two core events
at energies smaller than these of the alignment phenomenon.
As the energy increases,
the number of particles in the central part of the string, as well
as in its forward end, grows. When it is
high enough, some particles appear at the transitional angles.
Certainly not only the probability of production of a few high
energy particles at angles in between increases, but also their
energies grow, making them less sensitive to the cascading processes
in the later cascade development in the atmosphere.
Further calculation are needed in order to settle the details but the
fast rise of the rate of aligned events seams plausible taking into account
the common character of curving the relativistically rotating string.

Additionally, it is worthwhile mentioning that the discussed
mechanism leaves the question of the correlation
of energies and the positions of the observed energetic cores open.
The general absence of any clear correlation in experimental data
is difficult to explain by other models of alignment.

\section{Summary.}

We have proposed a mechanism which can be responsible for the alignment
of the very high energy interaction product observed in high altitude
cosmic ray experiments.

We postulate no ``new physics''. The unusual alignment of the creation
processwhich had previously been suggested as
extraordinary high momentum transfer processes or
new, exotic particles, could be strictly kinematical effect due to the
conservation laws. The conservation of the angular momentum in the
creation of fast rotating strings leads to its co-planar decay.
The problem of quantitative description of the hadronization of such object
needs detailed knowledge of the nature of the string -- chain, fireball
or jet.
Each of these words has its individual connotation and it has not yet been
decided
which (if any) describes the high energy particle production process.

The qualitative description of the relativistic rotating string could help to
explain the phenomena of binocular and aligned events seen in some cosmic ray
experiments.

\acknowledgments
It is a pleasure to tkank Prof. A.W. Wolfendale for a
carful reading of the manuscript.

\vspace{2cm}

\end{document}